\pdfoutput=1
\documentclass[sigconf,nonacm]{acmart}

\usepackage{tikz}
\usetikzlibrary{positioning,arrows.meta,decorations.pathreplacing,fit,calc}

\theoremstyle{acmdefinition}
\newtheorem{rulex}{Rule}

\definecolor{vnavy}{HTML}{1C154F}
\definecolor{vteal}{HTML}{0BB5BF}
\definecolor{vtealink}{HTML}{0A8A94}
\definecolor{vorange}{HTML}{F37025}

\newcommand{\step}[1]{\noindent\textsc{#1}}

\newcommand{\showURL}[1]{\unskip}

\begin{document}

\title{Mind the Intent Gap: A Methodology for Collaborative Agent-Driven Software Engineering}

\author{Nicolas Tacheny}
\orcid{0009-0006-0467-1719}
\affiliation{
  \institution{University of Mons (UMONS)}
  \city{Mons}
  \country{Belgium}
}
\affiliation{
  \institution{Ni2 Innovation Lab}
  \country{Belgium}
}
\email{nicolas.tacheny@{umons.ac.be,ni2.com}}

\author{Tom Mens}
\orcid{0000-0003-3636-5020}
\affiliation{
  \institution{University of Mons (UMONS)}
  \city{Mons}
  \country{Belgium}
}
\email{tom.mens@umons.ac.be}

\author{Saïd Mahmoudi}
\orcid{0000-0001-8272-9425}
\affiliation{
  \institution{University of Mons (UMONS)}
  \city{Mons}
  \country{Belgium}
}
\email{said.mahmoudi@umons.ac.be}


\begin{abstract}
AI agents have entered software teams, first as tools and now as team members. They produce in minutes what used to take days, yet the expected gain in delivery often fails to appear. We argue that they are entering processes designed for people, which quietly relied on people to cover three of their weaknesses. The team's intent was never fully written down, because colleagues filled the gaps. An AI agent fills them too, though by guessing, and drifts from what the team had in mind. The reasoning behind each deliverable was already lost from one step to the next, and AI agents now inflate and deflate the documents passed along. The reasoning people did alone was partly private and now moves into AI tools that keep it private by nature. We present CAIA, for Collaborative Agentic Intent Alignment, a methodology for software teams that brings agents into the way they work. A knowledge base describes the product that a project maintains. For each Change, a micro-team aligns on its intent before agents evolve the product and propose updates to the Knowledge Base. The proposed CAIA methodology emerged from an industrial need and is already in use at Ni2, with a platform built to support it. We present the methodology and outline our plans to evaluate it further.
\end{abstract}


\maketitle


\section{Introduction}
\label{sec:intro}

In recent years, AI agents, or agents for short, entered software development as assistive tools. They helped complete specific tasks defined by developers, such as writing a function, then a test, then a whole feature. These agents are now becoming full-fledged team members. While they were expected to speed up development, the evidence available to date does not fully confirm their added value. In a randomised trial run by METR in 2025, experienced open-source developers took 19\% longer to complete real issues when they could use AI tools~\cite{becker2025metr}. Open-source projects that adopted the Cursor assistant produced code much faster at first, then lost the gain as their code grew more complex~\cite{he2026cursor}. Each step gets faster while the work as a whole does not.

We hypothesise that one reason is that agents are entering processes that were designed for people. For each change, most teams organise their work as a chain: they decide what to make, then how to make it, then build it. At each stage of that chain, people implicitly did extra work. They filled in what the documents left out, explained their reasoning to the next team and shared the ideas they had worked out alone. Agents fill such gaps from what they learned in training, not from what the team has in mind.

Today, many software teams are in a phase of transition and want agents to join the team. We present the CAIA methodology, for Collaborative Agentic Intent Alignment, as a way to facilitate this transition. While vibe coding has been widely criticised because many humans just build with an agent and blindly trust its output, CAIA aims to bring more rigour, by providing a collaborative team methodology within the emerging field of agentic software engineering.

CAIA is based on the following main building blocks. A \emph{Knowledge Base} describes the \emph{Product} that a \emph{Project} maintains, for example a web application with all its code, tests and documentation. In each \emph{Change}, a small team of people aligns on its \emph{Intent} and writes it down. \emph{Agents} then evolve the product from that intent and propose how to update the knowledge base. CAIA is currently in use at Ni2, where it emerged from a real industrial need.

After presenting related work in Section~\ref{sec:related}, Section~\ref{sec:challenges} analyses the challenges that appear when agents join a team whose process was designed for people.
Section~\ref{sec:method} presents the main concepts of the intent-centred CAIA methodology, with the humans and agents that work on them. Section~\ref{sec:flow} describes the lifecycle of a Change, a flow of seven steps with explicit rules.
Finally, Section~\ref{sec:future} presents the current use of CAIA in a company setting, and outlines a plan to evaluate CAIA with students in a university setting.


\section{Related work}
\label{sec:related}

In 2025, vibe coding emerged as a new way of programming with coding agents. A developer describes in plain language what they want, the agent writes the code and the developer judges what comes back. Studying such sessions, Sarkar and Drosos~\cite{sarkar2025vibe} find that the developer's expertise shifts towards managing the context given to the agent. The conversation they observe involves one person and one agent. But in a team setting, how does that person know what the rest of the team has in mind, and how does the team learn what was decided?

Intent misalignment between team members is a long-standing challenge in software engineering. Bjarnason et al.~\cite{bjarnason2011gaps} traced how requirements slip through the gaps between the requirements engineers who decide what to build and the software engineers who build it. Agile methodologies answered with people, valuing individuals and interactions over processes and tools~\cite{beck2001manifesto}. Colleagues closed the gaps in conversation. When one person talks to an agent, those colleagues are no longer part of it.

Krentsel et al.~\cite{krentsel2026gaps} name a requirement gap between what stakeholders intend and what is written down, close to the gap at the heart of this paper, and narrow it with revision loops after the build. In spec-driven development, supported by tools such as GitHub Spec Kit and AWS Kiro, a structured specification drives the agent~\cite{agarwal2026specmine}. CAIA shares the idea of writing the intent down before the agent builds, yet differs in important ways: (1)~several disciplines agree on the intent, not a single developer; (2)~a knowledge base remembers what every change decided, not only one specification; (3)~rules govern who drives the agents and when the intent is frozen.

Storey~\cite{storey2026debt} describes how AI-driven development adds two debts to technical debt. Cognitive debt is the erosion of a team's shared understanding, and intent debt the absence of explicit rationale that humans and agents need to change a system safely. The CAIA methodology proposed in this paper keeps both debts low by writing down the intent of each change and sharing the reasoning behind it. Its knowledge base extends the architecture decision records proposed by Nygard~\cite{nygard2011adr} to the whole product and makes updating it part of every change. To the best of our knowledge, no existing methodology combines a collective alignment on intent, a single driver for the agents that change the product and a knowledge base that every change keeps up to date.


\definecolor{lead1}{HTML}{F4A261}
\definecolor{lead2}{HTML}{E9C46A}
\definecolor{lead3}{HTML}{6CA6E0}
\definecolor{lead4}{HTML}{B39DDB}
\begin{figure*}[t]
\centering
\resizebox{\linewidth}{!}{%
\begin{tikzpicture}[
  x=24mm, y=8mm,
  stepbox/.style={rounded corners=3pt, minimum width=19mm, minimum height=10mm,
                  font=\small\bfseries, text=white, align=center},
  flow/.style={-{Stealth[length=2.6mm]}, line width=1.2pt, vnavy!60},
  loop/.style={-{Stealth[length=2.2mm]}, line width=0.9pt, dashed, vnavy!60},
  looplabel/.style={font=\scriptsize\itshape, text=vnavy},
  phasebox/.style={rounded corners=5pt, line width=1.2pt},
  phase/.style={font=\normalsize\bfseries, anchor=south, text depth=0.25ex},
  brief/.style={draw, rounded corners=2pt, minimum height=5mm, minimum width=11mm,
                font=\scriptsize\bfseries, line width=0.8pt},
  source/.style={rounded corners=2pt, minimum height=4.5mm, font=\scriptsize\bfseries,
                 anchor=west, minimum width=171mm},
  agentname/.style={font=\scriptsize, anchor=west, align=left, inner sep=1pt},
  lead/.pic={
    \path[pic actions] (0,1.1mm) circle (0.75mm);
    \path[pic actions] (-1.2mm,-1mm) arc[start angle=180, end angle=0, radius=1.2mm] -- cycle;},
  master/.pic={
    \path[pic actions] (0,1.1mm) circle (0.75mm);
    \path[pic actions] (-1.2mm,-1mm) arc[start angle=180, end angle=0, radius=1.2mm] -- cycle;
    \path[pic actions] (-0.7mm,1.95mm) -- (-0.75mm,2.75mm) -- (-0.35mm,2.3mm) -- (0,2.85mm)
      -- (0.35mm,2.3mm) -- (0.75mm,2.75mm) -- (0.7mm,1.95mm) -- cycle;},
  bot/.pic={
    \path[pic actions, rounded corners=0.8mm] (-2.2mm,-1.6mm) rectangle (2.2mm,1.6mm);
    \fill[white] (-0.95mm,0.25mm) circle (0.55mm) (0.95mm,0.25mm) circle (0.55mm);
    \fill[white] (-0.9mm,-0.95mm) rectangle (0.9mm,-0.6mm);
    \draw[pic actions, line width=0.6pt] (0,1.6mm) -- (0,2.3mm);
    \path[pic actions] (0,2.55mm) circle (0.4mm);},
]
  \draw[phasebox, vnavy, fill=vnavy!4] (0.52,-1.33) rectangle (4.48,1.33);
  \node[phase, text=vnavy] at (2.5,1.33) {Alignment};
  \draw[phasebox, vtealink, fill=vteal!6] (4.52,-1.33) rectangle (6.48,1.33);
  \node[phase, text=vtealink] at (5.5,1.33) {Building};
  \draw[phasebox, vorange, fill=vorange!6] (6.52,-1.33) rectangle (7.48,1.33);
  \node[phase, text=vorange] at (7,1.33) {Consolidation};

  \foreach \i/\txt in {1/Describe, 2/Analyse, 3/Review, 4/Align}
    \node[stepbox, fill=vnavy] (s\i) at (\i,0) {\txt};
  \node[stepbox, fill=vteal] (s5) at (5,0) {Build};
  \node[stepbox, fill=vteal] (s6) at (6,0) {Validate};
  \node[stepbox, fill=vorange] (s7) at (7,0) {Update\\knowledge};
  \foreach \i [evaluate=\i as \j using int(\i+1)] in {1,...,6}
    \draw[flow] (s\i) -- (s\j);

  \draw[loop] (s4.north) to[out=165, in=15] node[looplabel, below=-1pt] {no-go} (s2.north);
  \draw[loop] (s6.north) to[out=160, in=20] node[looplabel, above=-1pt] {minor loop} (s5.north);
  \draw[loop] (4.75,1.33) to[out=90, in=90, looseness=1.2] node[looplabel, pos=0.75, left=2pt] {major loop} (4.25,1.33);

  \newcommand{\who}[2]{%
    \foreach \c/\r/\k [count=\j] in {#2} {
      \pgfmathsetmacro{\xo}{(\j-1)*0.12}
      \coordinate (w) at ($(#1)+(\xo,-1.12)$);
      \if\k f
        \if\r M\pic[fill=\c] at (w) {master};\else\pic[fill=\c] at (w) {lead};\fi
      \else
        \if\r M\pic[draw=\c, fill=white, line width=0.5pt] at (w) {master};%
        \else\pic[draw=\c, fill=white, line width=0.5pt] at (w) {lead};\fi
      \fi
    }}
  \who{1,0}{vnavy/M/f}
  \who{1.82,0}{lead1/L/f, lead2/L/f, lead3/L/f, lead4/L/f}
  \who{2.76,0}{vteal/M/f, lead1/L/f, lead2/L/f, lead3/L/f, lead4/L/f}
  \who{3.76,0}{vteal/M/f, lead1/L/f, lead2/L/f, lead3/L/f, lead4/L/f}
  \who{4.76,0}{vteal/M/f, lead1/L/m, lead2/L/m, lead3/L/m, lead4/L/m}
  \who{5.76,0}{vteal/M/f, lead1/L/f, lead2/L/f, lead3/L/f, lead4/L/f}
  \who{6.76,0}{vteal/M/f, lead1/L/m, lead2/L/m, lead3/L/m, lead4/L/m}

  \node[source, fill=vnavy!12, text=vnavy] (si) at (0.45,-1.75) {Stated intent of the Change};
  \node[source, fill=vorange!18, text=vorange!80!black] (kb) at (0.45,-3.65) {Knowledge Base};
  \foreach \x/\bx/\col/\name in {2.4/2.95/vnavy/{assistant\\agents},
                                 4.6/5.15/vteal/{build\\agent},
                                 6.6/7.15/vorange/{knowledge\\agent}} {
    \node[brief, draw=\col, fill=\col, text=white] (b) at (\x,-2.7) {brief};
    \draw[-{Stealth[length=2mm]}, line width=0.8pt, vnavy!60] (\x,-1.75) ++(0,-0.28) -- (b.north);
    \draw[-{Stealth[length=2mm]}, line width=0.8pt, vorange] (\x,-3.65) ++(0,0.28) -- (b.south);
    \pic[fill=\col, draw=\col] at (\bx,-2.72) {bot};
    \draw[-{Stealth[length=1.8mm]}, line width=0.8pt, \col] (b.east) -- (\bx-0.11,-2.7);
    \node[agentname, text=\col] at (\bx+0.1,-2.7) {\name};
  }

  \node[brief, draw=vtealink, fill=vteal!18, text=vtealink, minimum width=12mm] (art) at (6.06,-2.7) {Product};
  \draw[-{Stealth[length=2mm]}, line width=0.8pt, dashed, vnavy] (2.885,-2.52) -- (2.885,-2.03);
  \draw[-{Stealth[length=2mm]}, line width=0.8pt, vteal] (5.6,-2.7) -- (art.west);
  \draw[-{Stealth[length=2mm]}, line width=0.8pt, vorange] (7.15,-2.94) -- (7.15,-3.37);

  \draw[-{Stealth[length=2.6mm]}, line width=1.2pt, vorange]
    (s7.east) -- (7.82,0) -- (7.82,-3.65) -- (kb.east);
  \draw[-{Stealth[length=2.6mm]}, line width=1.2pt, vorange]
    (kb.west) -- (0.3,-3.65) -- node[font=\scriptsize\bfseries, rotate=90, above, text=vorange] {next Change} (0.3,0) -- (s1.west);

  \begin{scope}[shift={(0.55,2.15)}, every node/.style={font=\scriptsize, anchor=west}]
    \pic[fill=vnavy] at (0,-0.05) {master};  \node at (0.06,0) {Project Master};
    \pic[fill=vteal] at (0.9,-0.05) {master}; \node at (0.96,0) {Change Master};
    \foreach \c [count=\j] in {lead1, lead2, lead3, lead4}
      \pic[fill=\c] at (1.58+0.1*\j,-0.05) {lead};
    \node at (2.06,0) {Leads};
    \pic[draw=black!60, fill=white, line width=0.5pt] at (2.6,-0.05) {lead};
    \node at (2.66,0) {\itshape may join};
  \end{scope}
\end{tikzpicture}}
\caption{Schematic depiction of the life of a Change, its three phases, seven steps, actors involved  in each step and the brief its agents work from. The human actor that drives an agent draws its brief from the stated Intent and the Knowledge Base.}
\label{fig:flow}
\end{figure*}
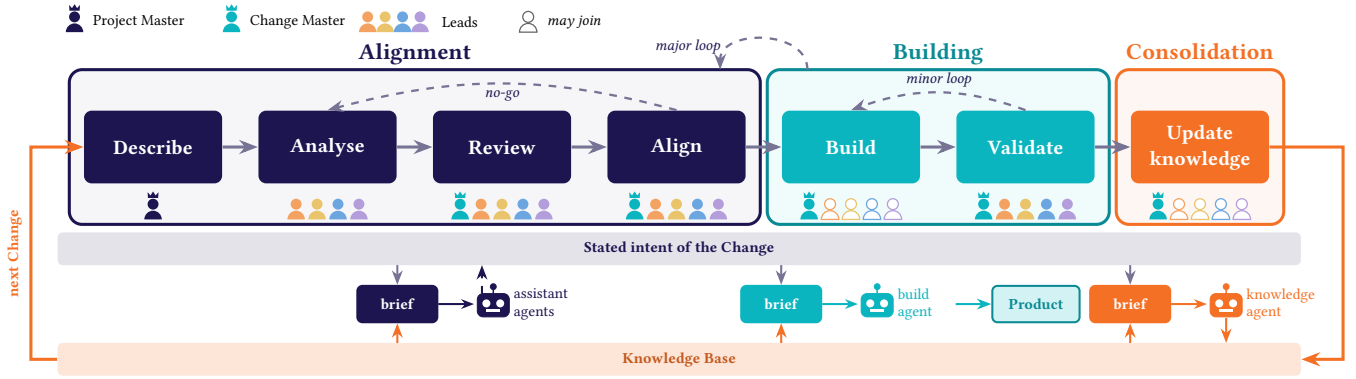

\section{Challenges when agents join the team}
\label{sec:challenges}

This section presents three challenges (C1 to C3) that may arise when agents join a project team whose process was designed primarily for people.
A team keeps changing what it maintains, and each change goes through three kinds of work: deciding on \emph{\textbf{what}} the change should achieve, deciding on \emph{\textbf{how}} to achieve this change, and carrying out the change (i.e., \emph{\textbf{build}} it). What this chain creates is the \textbf{Product} of the project, which is maintained and evolves over time because it has value of its own.
The product covers every deliverable, component and document needed to create and maintain it, from code to a website, each in several versions. One team may work on several projects, each with its own product.

\subsection*{C1. The intent gap}
A team about to create or evolve its product first agrees on what it wants, then writes it down for a colleague or for an  agent. What the team agrees on is the \textbf{Intent} of the change. Its human- and agent-readable form is the \textbf{Stated Intent}. The \textbf{Intent Gap} reflects what the intent holds and the stated intent does not declare.

\emph{What} and \emph{how} shape the Intent, while only \emph{build} changes the Product. The intent lives in discussions and in people's minds, so the two never match exactly. Before agents were used, the intent gap rarely hurt. A developer reading a thin ticket knew how the system was built, remembered previous discussions and could walk over and ask. Someone closed the gaps on the fly.

An agent, however, works only from what it is given for a task, which we call its \emph{LLM context}, and from what it learned in training. It can ask the person prompting it, but once the agent works for the team, that person is often not the one who knows. The answer may lie with a colleague from another discipline. Whatever is left unanswered, the agent invents or ignores while still returning a result that looks complete. Producing has become fast while closing the intent gap has not, and the gap has become the bottleneck.

\subsection*{C2. Reasoning lost between teams}

In a chained process each link explores options within its own boundary and passes on only its deliverable, such as a description of a need, a design or a plan. The reasoning that was made to produce it stays behind, and AI makes it worse. One link turns a three-line need into a forty-page document in one request, and the next person asks its personal agent for a five-bullet summary. The document is therefore inflated and deflated by agents without anybody reading it in full, and the original intent is lost between the two.

\subsection*{C3. Private reasoning}

Part of the reasoning behind any piece of work has always been private. Notes, sketches and ideas are rarely documented or shared and often remain in individuals' minds. With AI, people delegate a growing part of that reasoning to agents whose memory is private by nature, from chat histories to the memories agents keep about each user. When multiple persons collaborate on the same change, each person feeds their personal agent different LLM contexts and get as many versions of the same change. Merging them by hand reintroduces the coordination cost the agents were supposed to remove.


\section{The CAIA methodology}
\label{sec:method}

\subsection{Requirements}
Based on the challenges identified in Section~\ref{sec:challenges} we derive three main requirements for the proposed CAIA methodology:

\begin{itemize}
  \item[\textbf{R1}] The team members who hold the Intent of a change must explicitly state it, so that the intent gap is closed before any agent works from it.
  \item[\textbf{R2}] The reasoning behind a change travels with it to its end, and what was learned is kept for the next change.
  \item[\textbf{R3}] Several team members, each possibly assisted by their own agents, contribute to one shared stated intent, from which everything that changes the Product is driven.
\end{itemize}

\subsection{Concepts}

A \textbf{Project} evolves a \textbf{Product} one \textbf{Change} at a time, raising its value with each one. Each Change carries a \textbf{Stated Intent}, which grows as the project \textbf{Team} works on it. The team decides how large a Change is. It may add a capability, rework a part of the Product or fix a defect. The CAIA methodology does not predefine a fixed list of Change types. Alongside the Product, the Project maintains its evolving \textbf{Knowledge Base}, a body of knowledge that is complete enough to change the Product without relying on unverified assumptions. It covers what the Product is for, how it is organised, the conventions it follows and the decisions that shaped it, together with their reasons.

\subsection{Actors}

The project team is composed of both human and AI actors (i.e., agents), whose roles are described below.

\smallskip\noindent\emph{Human actors:} CAIA separates two kinds of contributions in a project, namely bringing expertise to the intent and operating the agents that act on it. We distinguish three human roles:
\begin{itemize}
  \item The \emph{Project Master} owns the project, builds the project team and assigns which of its members will act as \emph{Change Master}. The project master creates each Change and describes it.
  
  \item The \emph{Change Master} owns the Stated Intent of the Change they are responsible for, keeps the Change coherent and decides when it is ready to act on. The Change Master forms its own \emph{micro-team} composed of members from the project team.  A micro-team is small by design, and a large team splits into as many micro-teams as it has Changes in progress.

  \item A \emph{Lead} brings one discipline to the Change, such as business, user experience, architecture, development or quality. They contribute to the Stated Intent from that angle.
  The \emph{Change Master} assigns some Leads to analyse the Change and others to review it, ensuring that no Lead reviews their own analysis.
\end{itemize}

\smallskip\noindent\emph{AI actors:} An agent is built around an LLM, reasons on the context it receives, and acts or reacts on its environment through tools.  As challenge C1 revealed, an agent initially only knows its LLM context and its training. In CAIA, every agent receives the required context as a \textbf{Brief} (cf. Figure~\ref{fig:flow}), which gathers the relevant parts of the Stated Intent and of the Knowledge Base, along with instructions for the task.
CAIA distinguishes three kinds of agent:
\begin{itemize}
  \item An \emph{Assistant Agent} helps a human actor think. It can explore a topic, draft a text or challenge an argument. It never changes the Product or Knowledge Base.
  
  \item A \emph{Build Agent} evolves the Product from the Stated Intent.
  \item A \emph{Knowledge Agent} proposes updates to the Knowledge Base based on what the Change modified.
\end{itemize}

\begin{rulex}[One driver]
\label{rule:driver}
Only the Change Master operates the Build Agent and Knowledge Agent. Every human actor of a Change may use Assistant Agents at any step. Their output enters the Change only when a human writes it into the Stated Intent.
\end{rulex}

This rule ties humans and agents together,
and aims to ensure coherence. An agent's output is only as consistent as the brief it is given. If multiple human actors drove the agents, the briefs diverge and so do the results. One driver works from one Stated Intent. Whatever an actor works out alone with an assistant agent reaches the Product only through that intent, where the whole micro-team can read it (requirement R3). The quality of that intent, however, comes from the Leads, whose assistant agents help to express what they know.


\section{The life of a Change}
\label{sec:flow}

A Change goes through three main phases subdivided into seven steps (Figure~\ref{fig:flow}). First the micro-team aligns on the intent, next the build agent evolves the Product while the team checks it, and finally the knowledge agent proposes how the Knowledge Base should record the change.

\subsection{Alignment}

The first phase, called \emph{Alignment}, closes the intent gap before any build agent or knowledge agent acts. What colleagues used to fill in on the fly is now written down by the team members who hold it (requirement R1). Each step adds a block to the Stated Intent of the Change. As the Stated Intent grows block by block, any actor can draw a Brief from it at any moment, and use it together with the Knowledge Base to work on their block with the help of an assistant agent (Rule~\ref{rule:driver}). Since every discipline writes into the same Stated Intent, work never passes from one team to the next and the reasoning stays in its blocks, next to the decisions it led to (requirement R2).

The \emph{Alignment} phase is composed of four steps:

\step{Describe}: The Project Master formulates the Stated Intent as they first understand it, including why it matters and what is known to be in or out of scope.

\step{Analyse}: Each analyst Lead explores the Change from their discipline. For example, an architect identifies what the change will touch, and a quality Lead looks for what will be hard to verify.

\step{Review}: Each reviewer Lead, who did not participate in the analyses, reviews the Change independently and raises the risks and open questions seen from their discipline. The Change Master's own review looks at the Stated Intent as a whole and identifies the gaps that remain. Every review ends with a verdict, \emph{approved} or \emph{rejected}, and is written before the micro-team meets, so that no discipline is anchored by the others.

\step{Align}: The Change Master meets with all Leads to settle these gaps, if needed by sending the Change back to the \step{Analyse} step. The Change Master records the decisions and sets the final \emph{go} or \emph{no-go} verdict. A Lead's rejection carries no veto and is an invitation to discuss.

\subsection{Building}

Next comes the \emph{Building} phase, composed of two steps:

\step{Build}: Together with the Knowledge Base, the blocks written in the four Alignment steps form the \emph{build brief}. The Change Master gives it to the build agent and drives the agent until the Product has evolved as intended. If the Stated Intent left a decision open, the agent asks the Change Master instead of guessing. A \emph{solo build} suits a precise intent, and the Change Master builds alone. In a \emph{joint build} the team has left open parts of the solution on purpose, and the Leads discuss each choice as the agent proposes it. Either way, only the Change Master operates the agent (Rule~\ref{rule:driver}).

\step{Validate}: The micro-team checks the evolved Product against the Stated Intent. Small corrections send the Change back to the \emph{Build} step (minor loop), while a new understanding of the change sends it back to the \emph{Alignment} phase (major loop).

\subsection{Consolidation}

Once the Product is validated, the single-step \emph{Consolidation} phase keeps what was learned during the Change. At \step{Update Knowledge}, a \emph{knowledge brief} gathers what changed and the decisions taken, with their reasons, including those made during the build. The Change Master gives it to the knowledge agent, which proposes changes to the Knowledge Base without applying them. Once the Change Master, with the help of Leads when needed, has reviewed them, they update the Knowledge Base and the Change is finished. 
\begin{rulex}[A Change updates the Knowledge Base]
\label{rule:knowledge}
A Change that does not update the Knowledge Base is not finished.
\end{rulex}
Without this rule, the next Change would work from a Knowledge Base that no longer matches the Product.
Each Change must close a loop that raises the Product's value and makes the Knowledge Base ready for the next Change, whose briefs will read the decisions and their reasons (requirement R2).

\subsection{Rules of the flow}
\label{sec:rules}

The CAIA methodology enforces the aforementioned flow. Each Alignment step owns a block of the Stated Intent, which is either in progress or done, and a review is done once it is approved or rejected. A Change moves forward one step at a time and only when the current step is complete. The final verdict of \step{Align} must also be \emph{go}. No one can therefore reach the build without every analysis written and every reviewer having taken a position (R1). A Change may move back to any earlier step at any time, which distinguishes the methodology from a traditional waterfall process. The Stated Intent and the micro-team are frozen once \step{Build} starts, so the build brief cannot change underneath the agent.


\section{Current Use and Future Plans}
\label{sec:future}

CAIA emerged from a real need at Ni2, a company that builds software to manage digital infrastructure, from data centres to subsea cables. 
With about 50 employees in Belgium and Canada,
its geographically dispersed teams have long known intent gaps, since much of what a team intends never crosses the distance. The first author works at the Ni2 Innovation Lab and explored there how AI agents could help its teams work better.
This led to CAIA and to an automation platform built to support the methodology. The platform has been used since early 2026 by about ten company employees, for example to implement a new feature for a digital twin. At the time of writing, 19 Changes have been opened across 4 Projects, some of them still in progress. So far, as a first subjective evaluation, most team members describe the approach as positive.

For a more objective evaluation on an independent instantiation of CAIA, we are setting up a controlled experiment in a university setting, led by the second and third authors, in the context of a first course for engineering students in computer science and management and a second course for computer science master students. In each course, groups of students will receive the same software engineering problem, stated as a series of evolutions to carry out end to end on an existing piece of software. Some groups must follow CAIA, while the others can build freely with or without agents, with no imposed methodology. We will compare the groups on performance, such as the time each evolution takes, and on accuracy, such as how well each result matches the intended change. Our hypothesis is that the CAIA groups will deliver results closer to the intended change, with fewer corrections, in comparable time. 


\section{Conclusion}
\label{sec:conclusion}

AI agents have made each step of a team's work faster, but they join processes designed for people who filled the gaps in what was written, carried the reasoning along and shared what they had thought alone. In this paper, we proposed the CAIA methodology that makes that work part of every change. A micro-team aligns on the intent of each change before any AI agent builds, and the Knowledge Base keeps what the change leaves behind for the next one. Born from an industrial need, CAIA is in use at Ni2 today, and a planned controlled experiment at the University of Mons, Belgium will measure its effect. Language models were built on attention. The teams that build with them now need intent first.

\bibliographystyle{ACM-Reference-Format}
\bibliography{references}

\end{document}